\documentclass{ws-rv9x6}  
\usepackage{graphicx}
\providecommand{\openone}{\leavevmode\hbox{\small1\kern-3.0pt\normalsize1}}
\def\comment#1{}\def\labell#1{\label{#1}}

\begin{document}

\setcounter{chapter}{0}

\chapter{Homodyne tomography and the reconstruction of quantum states
  of light} 

\markboth{G. M. D'Ariano, L. Maccone, and M. F. Sacchi}{Homodyne
  tomography and the reconstruction of quantum states of light}

\author{Giacomo Mauro D'Ariano, Lorenzo Maccone, and Massimiliano Federico
Sacchi}

\address{QUIT - Quantum Information Theory Group, Dip. di Fisica ``A.
  Volta'', Universit\`a di Pavia, via A. Bassi 6, I-27100 Pavia
  \\ITALY}

\begin{abstract}  Quantum tomography is a procedure to determine the
  quantum state of a physical system, or equivalently, to estimate the
  expectation value of any operator.  It consists in appropriately
  averaging the outcomes of the measurement results of different
  observables, obtained on identical copies of the same system.
  Alternatively, it consists in maximizing an appropriate likelihood
  function defined on the same data. The procedure can be also used to
  completely characterize an unknown apparatus. Here we focus on the
  electromagnetic field, where the tomographic observables are
  obtained from homodyne detection.
\end{abstract}

\keywords{Quantum State Reconstruction, Quantum Tomography, Homodyne
  Detection, Maximum Likelihood, Quantum Calibration, Process
  Tomography.}

\section{Introduction}\labell{intro}     
The properties of each physical system are, by definition, completely
determined by its quantum state. Its mathematical description is given
in form of a density operator $\varrho$. Bohr's principle of
complementarity{\cite{complementarity}}, which is in many ways
connected with the uncertainty relations{\cite{comunc}}, forbids one
to recover the quantum state from a single physical system. In fact,
the precise knowledge of one property of the system implies that the
measurement outcomes of the complementary observables are all
equiprobable: the properties of a single system related to
complementary observables are simultaneously unknowable. Moreover, the
no-cloning principle{\cite{noclon}} precludes to obtain many copies of
a state starting from a single one, unless it is already
known\typeout{Notice that both the complementarity and the no-cloning
are not principles, and can be derived from the general postulates of
Quantum Mechanics. They are so called mainly for historical reasons.}.
Hence, complementarity and no-cloning prevent one to recover a
complete information starting from a single quantum system, i.e. to
recover its state. The only possibility is to recover it from multiple
copies of the system. [Notice that, if the multiple copies are not all
in the same quantum state, we will recover the mixed state of the
ensemble]. Given $N$ copies of a system, we can either perform a
collective measurement on all (or on subsets), or perform measurements
separately on each system and combine the measurement results at the
data analysis stage. Even though the former strategy would probably
increase the speed of the statistical convergence of the measured state to
the true one, it is quite impractical.  Tomography thus adopts the
latter strategy, which is the simplest to perform experimentally.

What is quantum tomography? It is the name under which all state
reconstruction techniques are denoted. It derives from the fact that
the first tomographic method (see Sec.~\ref{s:history}) employed the
same concepts of Radon-transform inversion we find in conventional
medical tomographic imaging. Since then, better methods have evolved
which eliminate the bias that the Radon-transform necessarily entails.
These fall into two main categories: the plain averaging method and
the maximum likelihood method. As will be seen in detail, the first
method requires a simple averaging of a function calculated on the $N$
measurement outcomes $x_n$ of the homodyne quadratures $X_{\phi_n}$.
Thus, the statistical error which affects the estimated quantity can
be easily evaluated through the variance of the data. The second
method, i.e. the maximum likelihood method, is based on the assumption
that the data we obtained is the most probable. Hence, we need to
search for the state that maximizes the probability of such data, i.e.
the state $\varrho$ for which $\prod_{i=1}^N\;_{\phi_n}\!\langle
x_n|\varrho|x_n\rangle_{\phi_n}$ is maximum, where $_{\phi_n}\!\langle
x_n|\varrho|x_n\rangle_{\phi_n}$ is the probability of obtaining the
result $x_n$ when measuring the quadrature $X_{\phi_n}$ (which has
eigenstates $|x\rangle_{\phi_n}$).

Their involved mathematical derivation has given these tomographic
techniques a false aura of being complicated procedures.  This is
totally unjustified: the reader only interested in applying the method
can simply skip all the mathematical details and proceed to
Sec.~\ref{s:dummies}, where we present only the end result, i.e. the
procedure needed in practice for a tomography experiment (the
experimental setup is, instead, given in Sec.~\ref{s:homdet}).

The chapter starts by introducing the method of homodyne tomography in
Sec.~\ref{s:homod}, along with the description of homodyne detectors,
noise deconvolution and adaptive techniques to reduce statistical
errors. Then, in Sec.~\ref{s:montec} we present the Monte Carlo
integration methods and the statistical error calculations that are
necessary for the plain averaging technique. In Sec.~\ref{s:maxlik},
the maximum likelihood methods are presented and analyzed. In
Sec.~\ref{s:dummies}, the step-by-step procedure to perform in
practice a tomography experiment is presented. In Sec.~\ref{s:device},
a tomographic method to calibrate (i.e.  completely characterize) an
unknown measurement device is presented.  Finally, in
Sec.~\ref{s:history}, a historical excursus on the development of
quantum tomography is briefly given.

\section{Homodyne tomography}\labell{s:homod}
The method of homodyne tomography is a direct application of the fact
that the displacements operators   ${\cal D}(\alpha)=e^{\alpha
a^\dag-\alpha^*a}$ are a complete orthonormal set for the linear space
of operators. Recalling that the scalar product in a space of
operators takes the Hilbert-Schmidt form $\langle
A|B\rangle=$Tr$[A^\dag B]$, this means that
\begin{eqnarray}
A=\int_{\mathbb C} \frac{d^2\alpha}{\pi}\;\mbox{Tr}[A\;{\cal
  D}^\dag(\alpha)]{\cal D}(\alpha)=\int_0^{\pi}\frac{d\phi}\pi
\int_{-\infty}^{+\infty}\!\!
dr\frac{|r|}4\mbox{Tr}[A\;e^{irX_\phi}]e^{-irX_\phi}
\;\labell{a},
\end{eqnarray}
where the polar variables $\alpha\equiv-ir\;e^{i\phi}/2$ were used in
the second equality.  Upon introducing the probability
$p(x,\phi)=\:_\phi\langle x|\varrho|x\rangle_\phi$ of obtaining
$x$  when measuring the quadrature $X_\phi =(a^\dag e^{i\phi }+a
e^{-i\phi})/2$, one obtains the tomographic formula 
\begin{eqnarray}
\langle A\rangle=\hbox{Tr}[A \varrho ]= \int_0^{\pi}\frac{d\phi}\pi
\int_{-\infty}^{+\infty} dx\;p(x,\phi)\;K_A(x,\phi)\;\labell{ba},
\end{eqnarray}
where
\begin{eqnarray}
K_A(x,\phi)\equiv\int_{-\infty}^{+\infty}dr\frac{|r|}4
\mbox{Tr}[A\;e^{ir(X_\phi-x)}]
\;\labell{bb},
\end{eqnarray}
defines the {\em kernel} of homodyne tomography. In the case of the
density matrix reconstruction in the Fock basis $|n\rangle$ (i.e. when
$A=|n\rangle\langle m|$), the kernel function is\cite{fermi}
\begin{eqnarray}
K_{A}(x,\phi)&=&2e^{i(m-n)\phi}\sqrt{\frac{m!}{n!}}
e^{-x^2}\sum_{j=m-n}^n\frac{(-1)^j}{j!}
\left(\begin{array}{c} n \cr m-j\end{array}\right)
\;\labell{fockbasis}
\\\nonumber&&\times
(2j+n-m+1)!\;\mbox{\sf Re}\Big[(-1)^{n-m}
{\cal D}_{-2(2j+n-m+2)}(-2ix)\Big]\;
,
\end{eqnarray}
where {\sf Re} denotes the real part and ${\cal D}_l(x)$ denotes the
parabolic cylinder function (which can be easily calculated through
its recursion formulas).

The multimode case is immediately obtained by observing that the
quadrature operators for different modes commute, so that for an
operator $A_M$ (acting on the Hilbert space of $M$ modes) we find
\begin{eqnarray}
&&\langle A_M\rangle=\int_0^{\pi}\frac{d\phi_1\cdots d\phi_M}{\pi^M}
\int_{-\infty}^{+\infty}
dx_1\cdots
dx_M\;p(x_1,\phi_1,\cdots,x_M,\phi_M)\nonumber\\
&&\qquad\qquad\times K_{A_M}(x_1,\phi_1,\cdots,x_M,\phi_M)
\;\labell{bc},
\end{eqnarray}
where $p(x_1,\phi_1,\cdots,x_M,\phi_M)$ is the joint probability of
obtaining the results $\{x_m\}$ when measuring the quadratures
$\{X_{\phi_m}\}$, and where
\begin{eqnarray}
K_{A_M}(x_1,\phi_1,\cdots)\equiv\int_{-\infty}^{+\infty}dr_1\cdots 
dr_M\prod_{m=1}^M\frac{|r_m|}4
\mbox{Tr}[A_M\;e^{ir_m(X_{\phi_m}-x_m)}]
\;\labell{bd1}.
\end{eqnarray}
However, such a simple generalization to multimode fields requires a
separate homodyne detector for each mode, which is unfeasible when the
modes of the field are not spatio-temporally separated. This is the
case, for example of pulsed fields, for which a general multimode
tomographic method is especially needed, because of the problem of
mode matching between the local oscillator and the detected fields
(determined by their relative spatio-temporal overlap), which produces
a dramatic reduction of the overall quantum efficiency. A general
method for multimode homodyne tomography can be found\cite{homtom}
that uses a {\em single} local oscillator that randomly scans all
possible linear combinations of incident modes.

\subsection{Homodyne Detection}\labell{s:homdet}
The balanced homodyne detector\cite{bilkentestim} measures the
quadratures $X_\phi\equiv(a^\dag e^{i\phi}+ae^{-i\phi})/2$. The
experimental setup is described in Fig.~\ref{f:homdet}. The
input-output transformations of the modes $a$ and $b$ that impinge
into a 50-50 beam-splitter are $c=(a+b)/\sqrt{2}$, $d=(a-b)/\sqrt{2}$
where $c$ and $d$ are the two beam-splitter output modes, each of
which impinge into a different photodetector.  The difference of the
two photocurrents is the homodyne detector's output, and thus is
proportional to $c^\dag c-d^\dag d=a^\dag b+b^\dag a$. In the strong
local oscillator limit, with mode $b$ in an excited coherent state
$|\beta\rangle$ $(|\beta|\gg 1)$, the expectation value of the output
is $I_H\propto\langle a^\dag\rangle\beta+\langle a\rangle\beta^*$
which is proportional to the expectation value of the quadrature
$X_\phi$, with $\phi$ the relative phase of the local oscillator.

\begin{figure}[ht]             
\centerline{\psfig{file=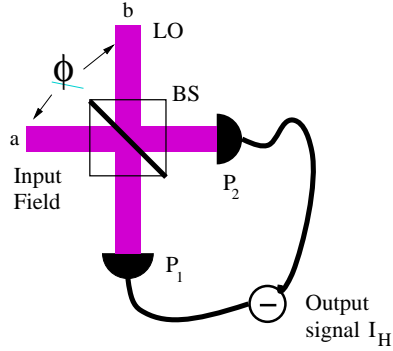,width=2.in}}
\vspace*{8pt}
\caption{Homodyne detector. The input signal (in mode $a$) is mixed
  by a 50-50 beam-splitter (BS) with a strong local oscillator (LO),
  which is coherent with the input field and is in a strong coherent
  state. The relative phase $\phi$ between the signal and the LO must
  be known and should be varied in $[0,\pi]$ with uniform probability.
  Two identical high efficiency linear photodetectors P$_1$ and P$_2$
  measure the field. The photocurrents are then accurately subtracted
  electronically yielding the output $I_H$.  Since the LO amplifies
  the weak quantum signals of the input, one can use high efficiency
  detectors that work only with strong signals.  \labell{f:homdet}}
\end{figure}

A detector with non-unit quantum efficiency $\eta$ is
equivalent\cite{mandelkelleykley} to a perfect $\eta=100\%$ detector,
preceded by a beam-splitter with transmissivity $\eta$.  Inserting two
beam-splitters in front of the two photodiodes of the homodyne scheme,
the modes $c$ and $d$ evolve as $c'=\sqrt{\eta}\;c+\sqrt{1-\eta}\;u$
and $d'=\sqrt{\eta}\;d+\sqrt{1-\eta}\;v$, where $u$ and $v$ are vacuum
noise modes. The homodyne output, is now proportional to $c'^\dag
c'-d'^\dag d'$, i.e. to
$L\equiv\eta\;(a^{\dag}b+b^{\dag}a)+(1-\eta)(u^{\dag}u-v^{\dag}v)
+\sqrt{{(1-\eta)\eta}/2}
[a(u^{\dag}-v^{\dag})+b(u^{\dag}+v^{\dag})+a^{\dag}(u-v)+b^{\dag}(u+v)]$.
As before, we take the limit $|\beta|\gg 1$ of strong pump in $b$, and
rescale the output difference photocurrent by $2|\beta|\eta$,
obtaining
\begin{eqnarray} I_H(\eta)=\lim_{|\beta|\to\infty}\frac{\langle
    L\rangle}{2|\beta|\eta}=\langle X_\phi\rangle+
\sqrt{\frac{1-\eta}{2\eta}}\langle u_\phi+v_\phi\rangle
\;\labell{outputhometa},
\end{eqnarray}
where the modes $u$ and $v$ are in the vacuum state. Since the quadrature
outcome for each vacuum state is Gaussian-distributed with variance
$1/4$, this means that the distribution of the noisy data are a 
convolution of the clean data with a Gaussian
of variance $\Delta_\eta^2=(1-\eta)/(4\eta)$, namely 
\begin{eqnarray}
p_\eta(x,\phi)=\frac 1{\sqrt{2\pi \Delta ^2_\eta}}
\int_{-\infty}^{+\infty} dx'e^{-(x-x')^2/(2\Delta^2_\eta)}\;p(x',\phi)
\;\labell{noisydata}.
\end{eqnarray}  
\subsection{Noise deconvolution}\labell{s:noise}
The data-analysis procedure can be modified to yield the result we
would obtain from perfect detectors, even though the data was
collected with noisy ones\cite{deconv}. In fact, depending on which
operator $A$ we consider and on the value of the quantum efficiency
$\eta$, the noise may be numerically deconvolved.  The output of the
noisy homodyne is distributed according to Eq.~(\ref{noisydata}), and
one can rewrite Eq.~(\ref{ba}) as follows
\begin{eqnarray} 
\langle A\rangle=\int_0^\pi \frac{d\phi}\pi \int_{-\infty} ^\infty 
dx\;p_\eta(x,\phi)\int_{-\infty} ^\infty dr\;\frac{|r|}4
\;e^{r^2\Delta_\eta^2/2}\;\mbox{Tr}[A\;e^{ir(X_\phi-x)}]
\;\labell{asb6},
\end{eqnarray}
where $p_\eta(x,\phi)$ is the probability of the noisy data. In the
case when all the integrals are convergent, the noise inversion can be
performed successfully.

It is clear the possibility of noise deconvolution depends on the
quantum efficiency of the detectors and the operator to be
estimated. For example, there is a bound $\eta>50\%$ for the
reconstruction of the density matrix in the Fock basis (i.e. for
$A=|n\rangle\langle m|$).  In fact, one can see that for $\eta<50\%$
Eq.~(\ref{asb6}) has an unbounded kernel. Notice that actual homodyne
detectors have efficiencies ranging between $70\%$ and $90\%$.

\subsection{Adaptive tomography}\labell{s:adap}
Adaptive tomography{\cite{adaptive}} exploits the existence of {\em
  null estimators} to reduce statistical errors. In fact, the addition
of a null estimator in the ideal case of infinite statistics does not
change the average of the data since, by definition, the mean value of
a null estimator is zero.  However, it can change the variance of the
data.  Thus, one can look for a procedure to reduce the variance by
adding suitable null functions.
\par In homodyne tomography null estimators are obtained as linear
combinations of the following operators
\begin{eqnarray}
{\cal N}_{k,n}(X_\varphi)=X_\varphi ^k\,e^{\pm
  i(k+2+2n)\varphi}\;,\qquad k,n\geq 0\;.
\end{eqnarray}
One can easily check that such functions have zero average over
$\varphi$, independently on $\varrho$. Hence, for every operator $A$
one actually has an equivalence class of infinitely many unbiased
estimators, which differ by a linear combination of functions   ${\cal
N}_{k,n}(X_\varphi )$. It is then possible to minimize the rms error
in the equivalence class by the least-squares method. This yields an
optimal estimator that is adapted to the particular set of
experimental data. Examples of simulations of the adaptive technique
that efficiently reduce statistical noise of homodyne tomographic
reconstructions can be found in Ref. {\cite{adaptive}}.

\section{Monte Carlo methods for tomography}\labell{s:montec}
In this section we will very briefly review the basics of the Monte
Carlo integration techniques that are needed and we show how to
evaluate the statistical error bars of the tomographically estimated
quantities.

A tomographic technique is based on an integral of the form
\begin{eqnarray}
F=\int_{-\infty}^{+\infty} dx\,p(x)\,f(x)
\;\label{integrpx},
\end{eqnarray}
where $p(x)$ is a probability. Since we have experimental outcomes
$\{x_n,\ n=1,\cdots\,N\}$ distributed according to the probability
$p(x)$, we sample the integral (\ref{integrpx}) using
\begin{eqnarray}
\int_{-\infty}^{+\infty} dx\,p(x)\,f(x)=\lim_{N\to\infty}\frac
1N\sum_{n=1}^Nf(x_n)
\;\label{cllimit}.
\end{eqnarray}
For finite $N$, the sum will be an unbiased estimator for the
integral, affected by statistical errors only (which can be made
arbitrarily small by increasing $N$).  The central limit theorem
guarantees that the finite sum $F_N=\sum_{n=1}^Nf(x_n)/N$ is a
statistical variable distributed as a Gaussian (for sufficiently high
$N$) with mean value $F$ and variance
\begin{eqnarray}
\sigma^2=\frac
1{N^2}\sum_{n=1}^N\lim_{M\to\infty}\frac 1M\sum_{j=1}^M\left[f(x_j)
\right]^2-F^2=\frac{\sigma^2(F)}N
\;\label{meanvare2}.
\end{eqnarray}
Hence, the tomographic estimated quantity converges with a statistical
error that decreases as $1/{\sqrt{N}}$. It can be estimated from the
data as
 \begin{eqnarray}
s^2(F_n)=\frac
1{N-1}\sum_{n=1}^N(F_n-m)^2\;.\labell{estvar}
\end{eqnarray}
[Remember that the factor $N-1$ in the variance denominator arises
from the fact that we are using the experimental estimated mean value
$m$ in place of the real one $F$.] The variance of the statistical
variable `mean $m$' is then given by $\sigma^2(m)=\sigma^2(F_N)/N$,
and thus the error bar on the mean $m$ estimated from the data is
given by
\begin{eqnarray}
\epsilon=\frac
1{\sqrt{N}}\;s(F_N)=\Big[\sum_{n=1}^N\frac{(F_n-m)^2}{N(N-1)}\Big]^{1/2}
\;\labell{esterrorbar}.
\end{eqnarray}
From the Gaussian integral one recovers the usual statistical
interpretation to the obtained results: the ``real'' value $F$ is to
be found in the interval $[m-\epsilon,m+\epsilon]$ with $\sim 68\%$
probability, in the interval $[m-2\epsilon,m+2\epsilon]$ with $\sim
95\%$ probability and in $[m-3\epsilon,m+3\epsilon]$ with $\sim$ unit
probability.

In order to test that the confidence intervals are estimated correctly
and that errors in the data analysis or systematic errors in the
experimental data do not undermine the final result, one may check the
$F_n$ distribution, to see if it actually is a Gaussian distribution.
This can be done by comparing a histogram of the data to a Gaussian,
or by using the $\chi^2$ test.  Notice that when we have very low
statistics it may be useful to use also bootstrapping techniques to
calculate the variance of the data.
\par For a more rigorous treatment of the statistical properties of
quantum tomography, and also some open statistical questions, see
Ref. \cite{gill}.

\section{Maximum likelihood tomography}\labell{s:maxlik}
The maximum likelihood tomography is based on the assumption that the
data obtained from the measurements is the most
likely\cite{maxlik}.  In contrast to the plain averaging
method presented above, the outcome is not a simple average of
functions of the data, but a Lagrange-multiplier maximization is
usually involved. The additional complexity introduced is compensated
by the fact that the results are statistically less noisy. Estimation
of operator expectation values is, however, indirect: one must first
estimate the state $\varrho$ and then calculate the expectation value
as Tr$[\varrho A]$.

Consider a known probability distribution $p_\gamma(x)$ parametrized
by a parameter $\gamma$ (which may also be a multidimensional
parameter).  We want to estimate the value of $\gamma$ from the data
set $\{x_1,\cdots,x_N\}$. The joint probability of obtaining such
data is given by the likelihood function
\begin{eqnarray}
{\cal L}(x_1,\cdots,x_N;\gamma)=\prod_{i=1}^Np_\gamma(x_k)
\;\labell{deflik}.
\end{eqnarray}
The maximum likelihood procedure consists essentially in finding the
$\gamma_0(x_1,\cdots,x_N)$ which maximizes the likelihood function
${\cal L}(x_1,\cdots,x_N;\gamma)$.  Equivalently, it may be convenient
to maximize its logarithm $\log{\cal L}(x_1,\cdots,x_N;\gamma)$, in
order to convert into a sum the product in Eq.~(\ref{deflik}). Usually, 
various constraints are known on the parameters $\gamma$, which can be
taken into account by performing a constrained maximization.  The
confidence interval for the estimated $\gamma_0$ can be evaluated from
the data using a bootstrapping technique: we can extract a rough
estimate of the probability distribution of the $\{x_i\}$ from the
data set, generate $M$ simulated sets of $N$ data points, and repeat
the procedure to obtain a set of $M$ parameters $\gamma_0^{(m)}$.
Their variance estimates the variance of the reconstruction. Moreover,
if a sufficiently large data set is present, we can attain the
Cramer-Rao bound $\sigma_\gamma^2\geqslant 1/NF_\gamma$, where
$F_\gamma$ is the Fisher information relative to $p_\gamma(x)$, i.e.
\begin{eqnarray}
F_\gamma\equiv\int dx\frac 1{p_\gamma(x)}
\left(\frac{\partial }{\partial\gamma}\;p_\gamma(x)\right)^2
\;\labell{deffisher}.
\end{eqnarray}
Since the Cramer-Rao bound is achieved only for the optimal
estimator\cite{cramer}, the maximum likelihood is among the best (i.e.
least statistically noisy) estimation procedures.

The maximum likelihood method can be extended to the quantum
domain\cite{maxlik}. The probability distribution of a measurement is
given by the Born rule as $p_i=$Tr$[\Pi_i\varrho]$ where $\{\Pi_i\}$
is the positive operator-valued measurement (POVM) that describes the
measurement. Thus we need to maximize the log-likelihood function
$L(\varrho)\equiv\sum_i\log\;$Tr$[\Pi_i\varrho]$ over the set of
density operators $\varrho$. In the case of finite Hilbert space,
$L(\varrho)$ is a concave function defined on a convex set of density
operators: its maximum is achieved on a single point or on a convex
subset. The main difficulty of this procedure consists in finding a
simple parameterization for the density matrix, that enforces both
the positivity and the normalization Tr$[\varrho]=1$.  The former is
guaranteed by requiring that $\varrho=T^\dag T$, the latter must be
taken into account through an appropriate Lagrange multiplier. In
order to employ the minimum number of parameters, it is sufficient to
consider $T$ as an upper complex triangular matrix with nonnegative
diagonal elements---so called Cholesky decomposition. 
This decomposition achieves minimal parameterization (up to the
normalization condition), as it requires $d^2$ real parameters for a
$d\times d$ Hermitian matrix. Thus, in practice we need to maximize
the operator $L_\lambda[\varrho]\equiv\sum_i\log$Tr$[\Pi_iT^\dag
T]-\lambda[T^\dag T]$, where $\lambda$ is a Lagrange multiplier that
accounts for the normalization. By expressing $\varrho$ in terms of
its eigenstates as $\varrho=\sum_my^2_m|\psi_m\rangle\langle\psi_m|$,
the condition for the maximum, $\partial L_\lambda/\partial y_m=0$,
becomes
\begin{eqnarray}
\sum_i \{ y_m\langle\psi_m|\Pi_i|\psi_m\rangle/
\mbox{Tr}[\varrho\Pi_i]\} - \lambda\;y_m=0\qquad \forall m
\;\labell{lagr}.
\end{eqnarray}
Multiplying both members by $y_m$ and summing over $m$, through the
Born rule and the normalization of $\varrho$, we find that $\lambda$
is equal to the number of measurements employed. Thus, we are left
with the problem of finding the maximum of the $d^2$-parameter
function $L_{\lambda=N}[\varrho=T^\dag T]$, which can be tackled with
conventional numerical techniques such as expectation-maximization or
downhill simplex\cite{maxlik}. By using the ML method only small
samples of data are required for a precise determination, even in the
presence of low quantum efficiency at the detectors.  However, we want
to emphasize that such method is not always the optimal solution of
the tomographic problem, since it suffers from some major limitations.
Besides being biased due to the Hilbert space truncation---even though
the bias can be very small if, from other methods, we know where to
truncate---it cannot be generalized to the estimation of any ensemble
average, but just of a set of parameters from which the density matrix
depends. In addition, for the multi-mode case, the method has
exponential complexity versus the number of modes.

\section{Tomography for dummies}\labell{s:dummies}
In this section we just give the step-by-step procedure to implement a
tomography experiment, employing all the results obtained in the
previous sections.\par\noindent$\bullet$~Plain averaging
method:\vspace{-.2cm}\begin{enumerate}\item Calculate the Kernel
function $K_A$ for the operator $A$ whose expectation value we want to
estimate through Eq.~(\ref{bb}). For example, to estimate the density
matrix in the Fock basis, we need the $K_A$ defined in
Eq.~(\ref{fockbasis}).\item The experimental apparatus, described in
Sec.~\ref{s:homdet}, yields a set of $N$ data points $\{\phi_n,x_n\}$
: each datum is composed by the quadrature phase $\phi_n$ that was
measured and by the corresponding measurement result $x_n$
.\item\labell{pp1} Evaluate $\frac 1 N \sum _ n K_A(x _n, \phi _n
)$. In the limit $N\to\infty$ this average yields the expectation
value $\langle A\rangle$ we are looking for.\item\labell{pp2} For
finite $N$, we can estimate the purely statistical error on the result
through Eq.~(\ref{esterrorbar}), replacing $m$ with the average
obtained at the previous point and $F_n$ with the $n$th Kernel
function evaluation,
$K_A[x_n,X_{\phi_n})]$.\end{enumerate}\vspace{-.2cm} Further data
massaging is also possible: we can employ adaptive tomography to
reduce the statistical noise (see Sec~\ref{s:adap}). Moreover, we can
remove the detector noise due to homodyne measurements with non unit
quantum efficiency $\eta$, as long as $\eta>1/2$ (see Sec.~\ref{s:noise}).
\vspace{.2cm}\par\noindent$\bullet$~Maximum likelihood
method:\vspace{-.2cm}\begin{enumerate}\item Parametrize the unknown
  quantum state through the upper triangular $d\times d$
  matrix $T$ as $\varrho=T^\dag T$. \item Use the same experimental
  apparatus (homodyne detection) to obtain $N$   data points
$\{\phi_n,x_n\}$. Calculate the log likelihood function on the
experimental data as $\log\sum_{n=1}^N\;_{\phi_n}\!\langle x_n|T^\dag
T|x_n\rangle_{\phi_n}$.\item Numerically maximize this quantity over
the $d^2$ parameters of $T$ with the additional constrain Tr$[T^\dag
T]=1$. This maximum is achieved on our best estimate for the state
$\varrho=T^\dag T$.\item The confidence intervals for our estimation
can be obtained using bootstrapping techniques, or employing the
Cramer-Rao bound of
Eq.~(\ref{deffisher}).\end{enumerate}

\section{Quantum calibration of measurement devices}\labell{s:device}
In this section we review the method to measure the POVM of an unknown
measurement apparatus presented in Ref.~\cite{calib}. The method is
based on analyzing the correlations in measurements on a bipartite
system: one of the two parts is fed into the unknown apparatus~A,
while the other is measured with a known set~B of detectors that
measures a quorum of observables (see Fig.~\ref{f:calib}). As will be
shown in the following, there is ample freedom in the choice of both
the input bipartite states and the set of observables. The procedure
is repeated many times and the joint measurement outcomes are analyzed
using the tomographic algorithms described above, which (in the limit
of infinite input data) yield the POVM of the unknown apparatus.  For
finite data, the reconstructed POVM will be affected only by
statistical errors which can be easily estimated. For the sake of
illustration, a Monte-Carlo simulation of the procedure is given at
the end of this section. It aptly illustrates the advantage of using
maximum likelihood techniques over plain averaging: the maximum
likelihood reconstruction is significantly less noisy.

\begin{figure}[ht]             
\centerline{\psfig{file=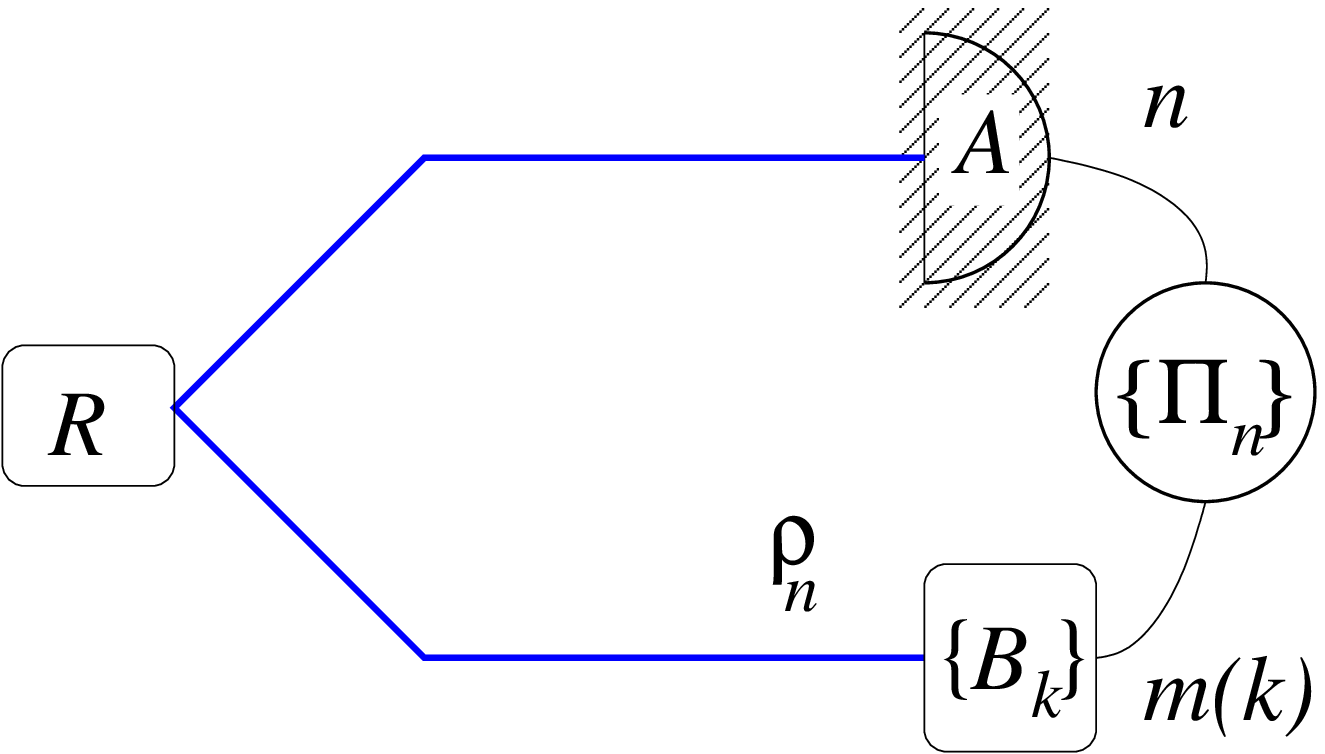,width=2.in}
\hspace{1cm}\psfig{file=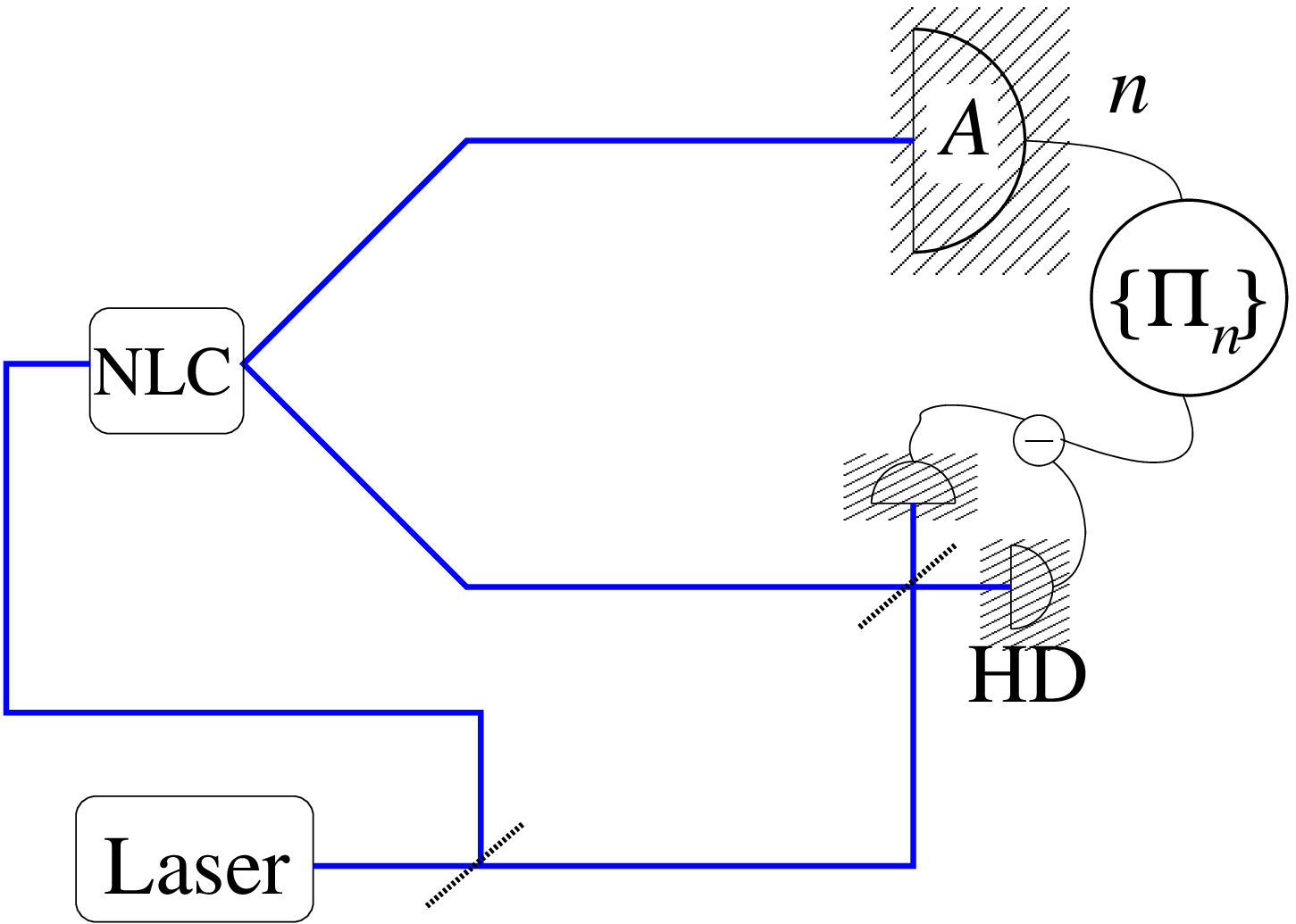,width=2.in}}
\vspace*{8pt}
\caption{(Left)~Experimental setup to determine the POVM of
  the unknown measurement apparatus~A: one part of the bipartite input
  state~$R$ is sent to the apparatus~A which yields the measurement
  result~$n$; the other part (with quantum state $\varrho_n$) is sent to
  the known detector~B which performs a projective measurement of an
  observable $B_k$ from the complete set $\{B_k\}$ yielding the result
  $m(k)$. The joint measurement results are processed using a
  tomographic algorithm to obtain the POVM $\{\Pi_n\}$ of~A.
  (Right)~Example of application of the scheme to the radiation field.
  The bipartite state $R$ is generated via a non-linear crystal
  through spontaneous parametric down-conversion. The tomographer B
  is, in this case, a homodyne detector (HD) which measures the
  quadratures, a complete set of observables.  \labell{f:calib}}
\end{figure}

The following simple example illustrates how the procedure works.
Suppose we want to evaluate the POVM of a von Neumann measurement of
the observable $O$ which acts on a $d$-dimensional Hilbert space
${\cal H}_A$ and has spectral decomposition
$\sum_no_n|o_n\rangle\langle o_n|$. We can use the maximally entangled
input state $|\Psi\rangle=\sum_{i=1}^d|i\rangle|i\rangle/\sqrt{d}$,
which lives in the space ${\cal H}_A\otimes{\cal H}_B$. In fact, this
state can be also written as
\begin{eqnarray}
|\Psi\rangle=\frac 1{\sqrt{d}}\sum_{i,j=1}^d\Big(|o_j\rangle\langle
o_j|\otimes\openone\Big)|i\rangle|i\rangle=
\frac 1{\sqrt{d}}\sum_{j=1}^d|o_j\rangle|o_j^*\rangle,
\labell{redux1}
\end{eqnarray}
where $*$ denotes complex conjugation with respect to the basis
$|i\rangle$.  It is obvious from Eq.~(\ref{redux1}) that the outcome
$o_n$ at detector A (corresponding to the state $|o_n\rangle$ in
${\cal H}_A$) means that the state $\varrho_n=|o_n^*\rangle\langle
o_n^*|$ in ${\cal H}_B$ impinges in detector~B.  The POVM can be
recovered using tomographical state reconstruction at B, since in this
simple case $\Pi_n=\varrho^*_n$.

It is not difficult to generalize the above example to arbitrary POVMs
and measurement procedures.  Let the unknown apparatus A be described
by the POVM $\{\Pi_n\}$ we want to estimate, and let the apparatus B
measure the quorum observables $O_k$ described by the von Neumann
projections $\{|k_m\rangle\langle k_m|\}$ (with $\{|k_m\rangle\}$
basis for all $k$).  From the Born statistical formula we can derive
the state that impinges into the known detector B if the unknown
detector A gave result $n$ for the measurement on the initial
bipartite state $R$, as
\begin{eqnarray}
\varrho_n=\frac{\mbox{Tr}_1[(\Pi_n\otimes\openone)R]}
{\mbox{Tr}[(\Pi_n\otimes\openone)R]}
\;\labell{pp1bis}.
\end{eqnarray}
It describes the state reduction at B stemming from a measurement at A
with outcome~$n$. The denominator is the probability $p(n)$ of
obtaining the result~$n$ at B. The state $\varrho_n$ contains some
information on the POVM element $\Pi_n$. It can be recovered by
introducing the map   ${\cal
R}(X)\equiv\mbox{Tr}_1[(X\otimes\openone)R]$, so that
Eq.~(\ref{pp1bis}) rewrites as $\varrho_n={\cal R}[\Pi_n/p(n)]$. This
implies that the POVM can be recovered as   $\Pi_n=p(n){\cal
R}^{-1}(\varrho_n)$, where the map $\cal R$ depends only on the input
state $R$: the input state $R$ allows the POVM reconstruction if the
inverse map ${\cal R}^{-1}$ exists.  This condition can be cast in a
more transparent form by rewriting the map $\cal R$ in a
multiplicative form via isomorphism between operators on   ${\cal
H}\otimes{\cal H}$ and maps\cite{qop1}. We can obtain an operator of
this form by considering $S=R^{T_1}$, i.e. the partial transposition
on the first space of the input state $R$.  In fact, taking
two operators $X$ and $Y$ such that $Y={\cal R}(X)$, we see that
\begin{eqnarray}
Y_{il}=\sum_{jk}X_{jk}\langle i|{\cal R}\Big(|j\rangle\langle
k|\Big)|l\rangle=\sum_{jk}X_{jk}(R^{T_1})_{jk,il}
\;\labell{defs},
\end{eqnarray}
where $Y_{il}=\langle i|Y|l\rangle$, $X_{jk}=\langle j|X|k\rangle$,
and $(R^{T_1})_{jk,il}=\langle j|\langle i|R^{T_1}|k\rangle|l\rangle$,
the set $\{|n\rangle\}$ being a basis in $\cal H$. In matrix notation
(considering $jk$ and $il$ as collective indexes), Eq.~(\ref{defs})
rewrites as $Y=SX$.  It follows immediately that the map $\cal R$ is
invertible if $S^{-1}$ exists so that $X=S^{-1}Y$. In this case we say
that the input state $R$ is faithful\cite{qop1}.  Since invertibility
is a condition satisfied by a dense set of operators, the set of input
states $R$ that allow the POVM reconstruction is also dense, i.e.
almost any bipartite state will do. In particular, all Gaussian
bipartite states---with the trivial exception of product states---are
faithful \cite{gfaith}. To recapitulate: in order to check
whether the state $\varrho_n$ allows to obtain the POVM (i.e.  whether
the input state $R$ is faithful) we must verify that the operator
$(R^{T_1})_{jk,il}$ is invertible when $jk$ and $il$ are considered as
collective indexes. As an illustration of this check, take the simple
example given above: the state
$|\Psi\rangle=\sum_i|ii\rangle/\sqrt{d}$ is faithful since
$|\Psi\rangle\langle\Psi|^{T_1}=\sum_{ij}|ji\rangle\langle ij|/d$ is
invertible: it is a multiple of the swap operator
$E\equiv\sum_{ij}|ji\rangle\langle ij|$.

To recover $\varrho_n$ from the measurements at B (and hence the POVM if
the input $R$ is faithful), we can use the quantum tomographic
techniques described in the previous sections. If we employ the plain
averaging technique, we may recover the density matrix elements
$\varrho_{ij}$ in some basis and then calculate the POVM using the
inverse map ${\cal R}^{-1}$, as
\begin{eqnarray}
\langle j|\Pi_n|k\rangle=p(n)\sum_{il}\varrho^{(n)}_{il}\:(R^{T_1})^{-1}_{jk,il},\nonumber 
&&\;\labell{melpovm}
\end{eqnarray}
where the inverse of $R^{T_1}$ must be calculated considering $jk$ and
$il$ as collective indexes. On the other hand, if we employ maximum
likelihood we may directly maximize the probability of acquiring the
data we obtained from the measurements\cite{maxlik}, i.e. the joint
probability $p_k(n,m)=$Tr$[(\Pi_n\otimes |k_m\rangle\langle k_m|)R]$.
Equivalently, one can maximize the logarithm of this quantity and
consider simultaneously all the $N$ joint measurement outcomes
$\{n_1,m_1\},\cdots,\{n_N,m_N\}$ of the quorum operators $O_{k^{(i)}}$
at detector A and of the unknown detector B. Thus, the POVM
$\{\Pi_n\}$ is the one that maximizes the quantity
\begin{eqnarray}
{\cal L}(\{\Pi_n\})\equiv\sum_{i=1}^N\log\mbox{Tr}\Big[(\Pi_{n_i}\otimes
|k^{(i)}_{m_i}\rangle\langle k^{(i)}_{m_i}|)R\Big] 
\;\labell{mlik},
\end{eqnarray}
with the additional constraints $\Pi_n\geqslant 0$ and
$\sum_n\Pi_n=\openone$.  Other prior knowledge on the quantities to be
estimated can be easily introduced adding further constraints to the
maximization. Also in this case it is possible to take into account a
known source of noise at the detector B: if we replace the term
$|k^{(i)}_{m_i}\rangle\langle k^{(i)}_{m_i}|$ in Eq.~(\ref{mlik}) with
the noise-evolved ${\cal N}(|k^{(i)}_{m_i}\rangle\langle
k^{(i)}_{m_i}|)$, then the maximization yields the POVM that maximizes
the {\it noisy} measurement results.

For the sake of illustration, we give a Monte-Carlo simulation of the
calibration procedure in which we recover the POVM of a simple
inefficient photodetector\cite{calib}. An inefficient photodetector is
aptly modeled by a perfect photodetector (which is a device which
measures the observable ``number of photons'' $a^\dag
a=\sum_nn|n\rangle\langle n|$), preceded by a beam-splitter with a
transmissivity equal to the quantum efficiency $\eta$ of the detector.
Possible dark counts can be considered by feeding the other
beam-splitter port with a thermal state with $\bar n$ average photons.
In this case, the theoretical POVM is given by
\begin{eqnarray}
&&\Pi_n=\sum_{p=0}^\infty|p\rangle\langle p|
\;\labell{povmd}
\\&&\nonumber\times\sum_{k=0}^\infty\sum_{j=0}^{\min(p,k+n)}
\left(\begin{array}{c} p \cr j\end{array}\right)
\left(\begin{array}{c} -n-1 \cr k\end{array}\right)
\left(\begin{array}{c} k+n \cr j\end{array}\right)
\eta^j(1-\eta)^{k+n-j}\;\bar n^{k+n-j}
\;.
\end{eqnarray}
Since this POVM is diagonal in the Fock basis, we can limit the
reconstruction to the diagonal elements.  As input state $R$ we employ
a twin beam state $|TB\rangle$, i.e. the result of spontaneous
parametric down-conversion:
\begin{eqnarray}
|TB\rangle\equiv\sqrt{1-|\xi|^2}\sum_m\xi^m|m\rangle_a|m\rangle_b
\;\labell{spdc},
\end{eqnarray}
where $\xi$ is the parametric amplifier gain and $|m\rangle_a$ and
$|m\rangle_b$ are Fock states of the modes $a$ and $b$ that impinge in
the detectors A and B respectively. This is a faithful state since
$|TB\rangle\langle TB|^{T_1}=(1-|\xi|^2)E\:\xi^{a^\dag
  a}\otimes{\xi^*}^{b^\dag b}$ (where $E$ is the swap operator) is
invertible. The photon counter measures the mode $a$ at
position~A, while homodyne detection with quantum efficiency $\eta_h$
measures the mode $b$ at position~B acting as tomographer (see
Fig.~\ref{f:calib}). Since only the diagonal part of the POVM is
needed, we can use a homodyne detector with uniformly distributed
local oscillator phase.  [A phase-controlled homodyne detector would
allow to recover also the off-diagonal elements of the POVM, ensuring
a complete characterization of the device.]

In Figs.~\ref{f:tomog} and~\ref{f:maxlik} we present the results of
the POVM reconstruction deriving from the two tomographic methods
described above (simple averaging and maximum likelihood, 
respectively).  The convergence of the maximum likelihood procedure is
assured since the likelihood functional $\cal L$ is convex over the
space of diagonal POVMs.  However, the convergence speed can become
very slow: in the simulation of Fig.~\ref{f:maxlik} a mixture of
sequential quadratic programming (to perform the constrained
maximization) and expectation-maximization techniques 
were employed. From the graphs it is evident that the maximum
likelihood estimation is statistically more efficient since it needs
much less experimental data than tomography. This is a general
characteristic of this method, since if the optimal estimator (i.e.
the one achieving the Cramer-Rao bound) exists, then it is equal to
the maximum likelihood estimator\cite{maxlik}. An added bonus,
evident from Eq.~(\ref{mlik}), is that the maximum likelihood recovers
all the POVM elements at the same time additionally increasing the
statistical efficiency. On the other hand, the tomographic
reconstruction is completely unbiased: no previous information on the
quantity to be recovered is introduced.

This simulated experiment uses realistic parameters and is feasible in
the lab with currently available technology\cite{homodyne}.  The
major experimental challenge lies in the phase matching of the
detectors, i.e. in ensuring that the modes detected at A and B
actually correspond to the modes $a$ and $b$ of the state
$|TB\rangle$.  

\begin{figure}[h!]
\begin{center}
\epsfxsize=.45
\hsize\leavevmode\epsffile{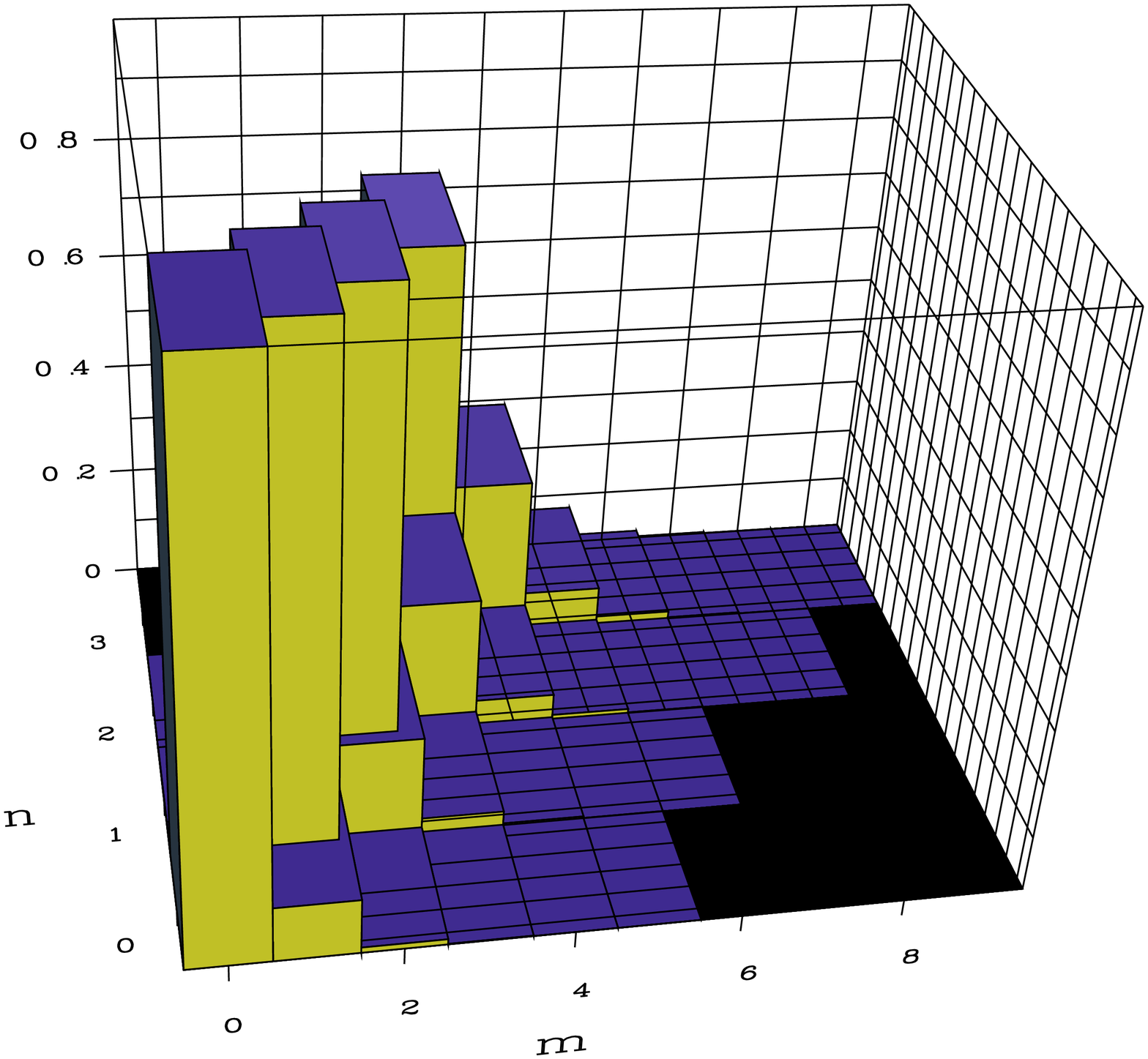}\hspace{1cm}
\epsfxsize=.45
\hsize\leavevmode\epsffile{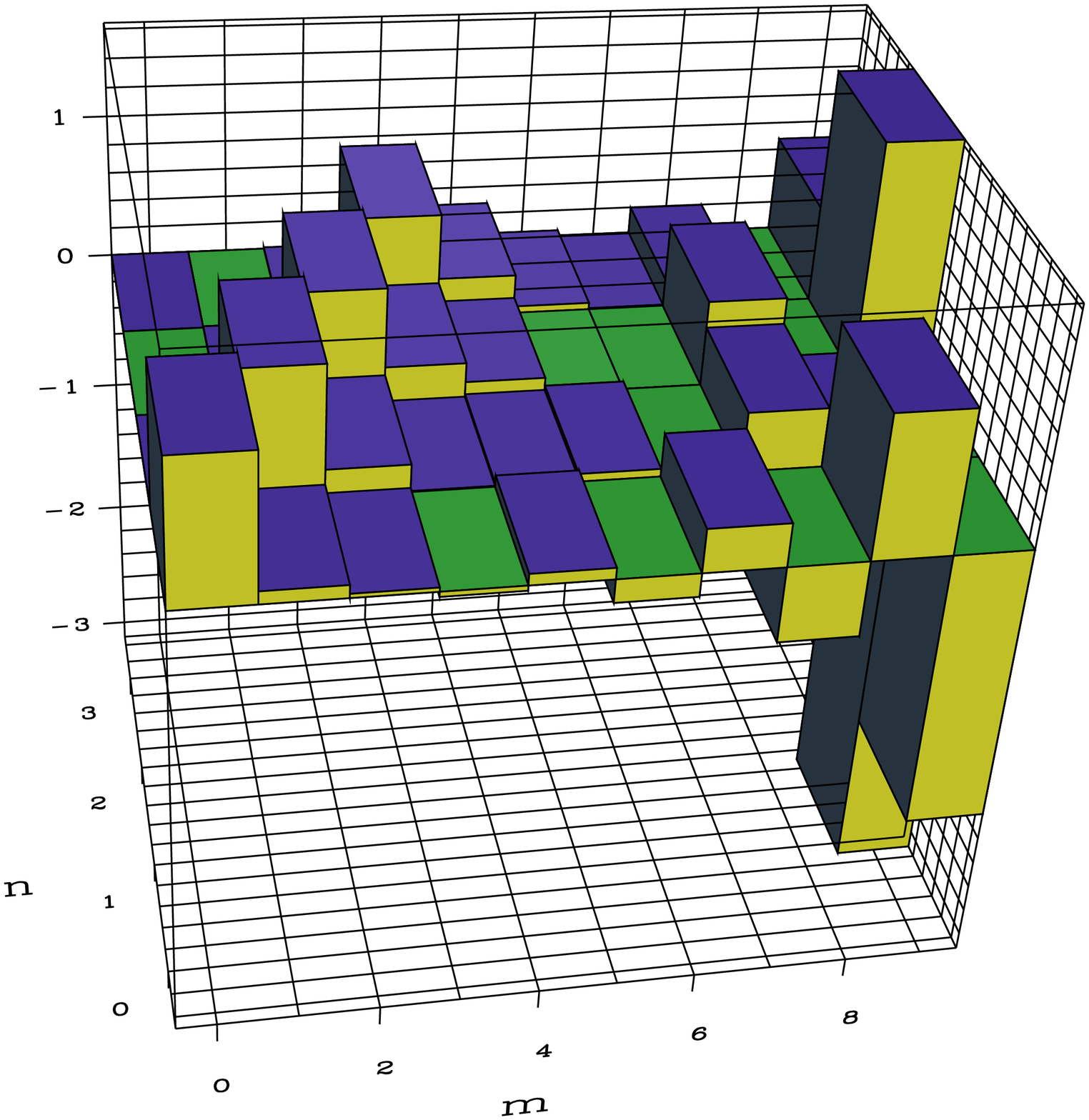}
\end{center}
\begin{center}
\epsfxsize=.5
\hsize\leavevmode\epsffile{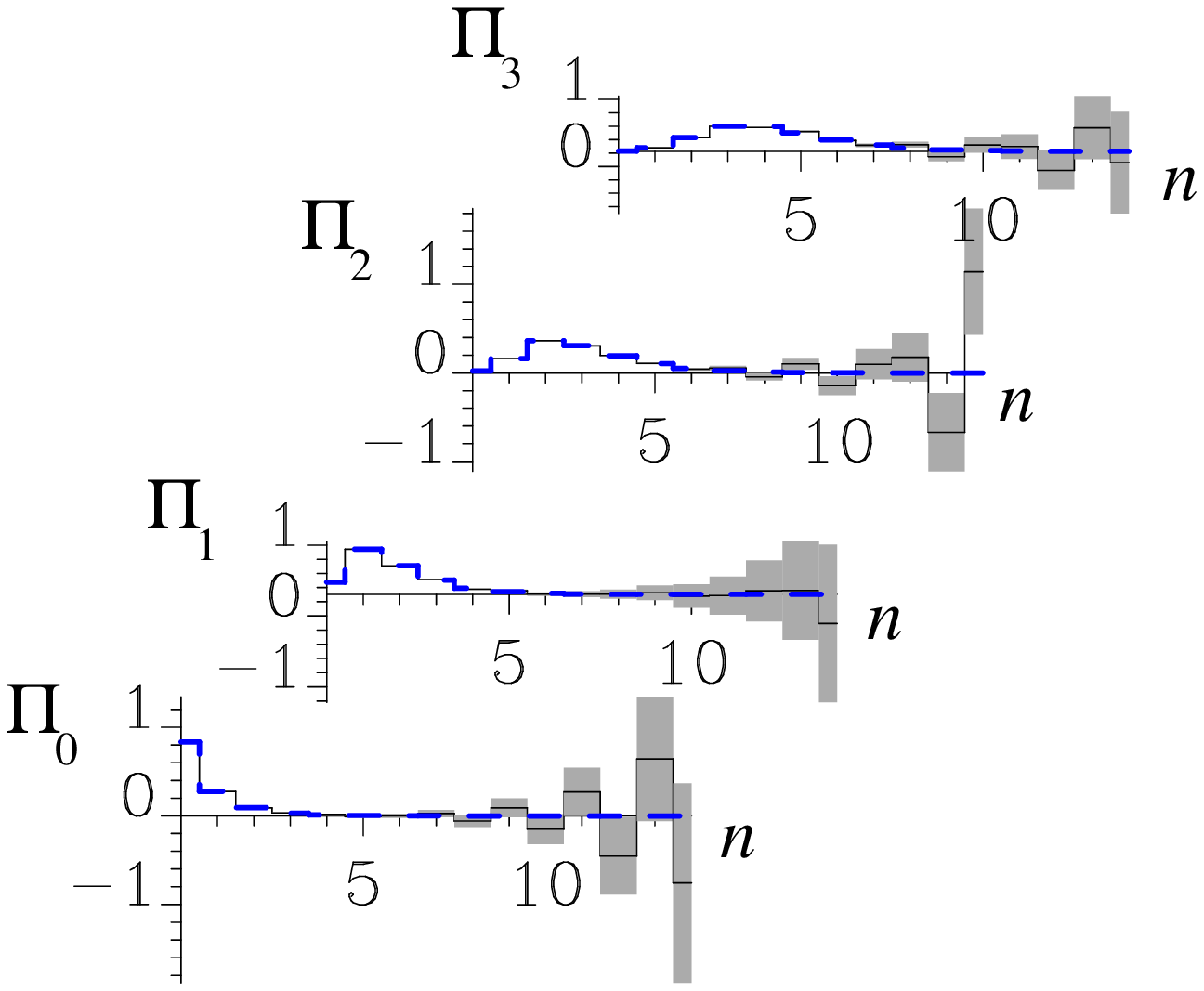}
\end{center}
\caption{(Above left) Theoretical value of the diagonals 
  of the POVM elements $\langle m|\Pi_n|m\rangle$ of the inefficient
  photodetector described by Eq.~(\ref{povmd}), with parameters $\bar
  n=1$, $\eta=80\%$. (Above right) Simulated reconstruction of the
  same quantity. The data are simulated as coming from an input
  twin-beam state $|TB\rangle$ with $\xi=0.88$, and as being detected
  from a phase insensitive homodyne detector with quantum efficiency
  $\eta_h=90\%$. Here $5\times 10^6$ simulated homodyne measurements
  are employed.  (Below)~The same data is plotted separately for each
  POVM element to emphasize the error bars.  They are obtained from
  the root-mean-square of the recovered POVM matrix elements.  (The
  theoretical value is plotted as the thick dashed line.) Plain
  tomographic averaging with noise deconvolution has been employed
  here, since the noise map of inefficient homodyne detection can be
  inverted for $\eta_h>50\%$.}  \labell{f:tomog}\end{figure}

\begin{figure}[htb]
\begin{center}
\epsfxsize=.45
\hsize\leavevmode\epsffile{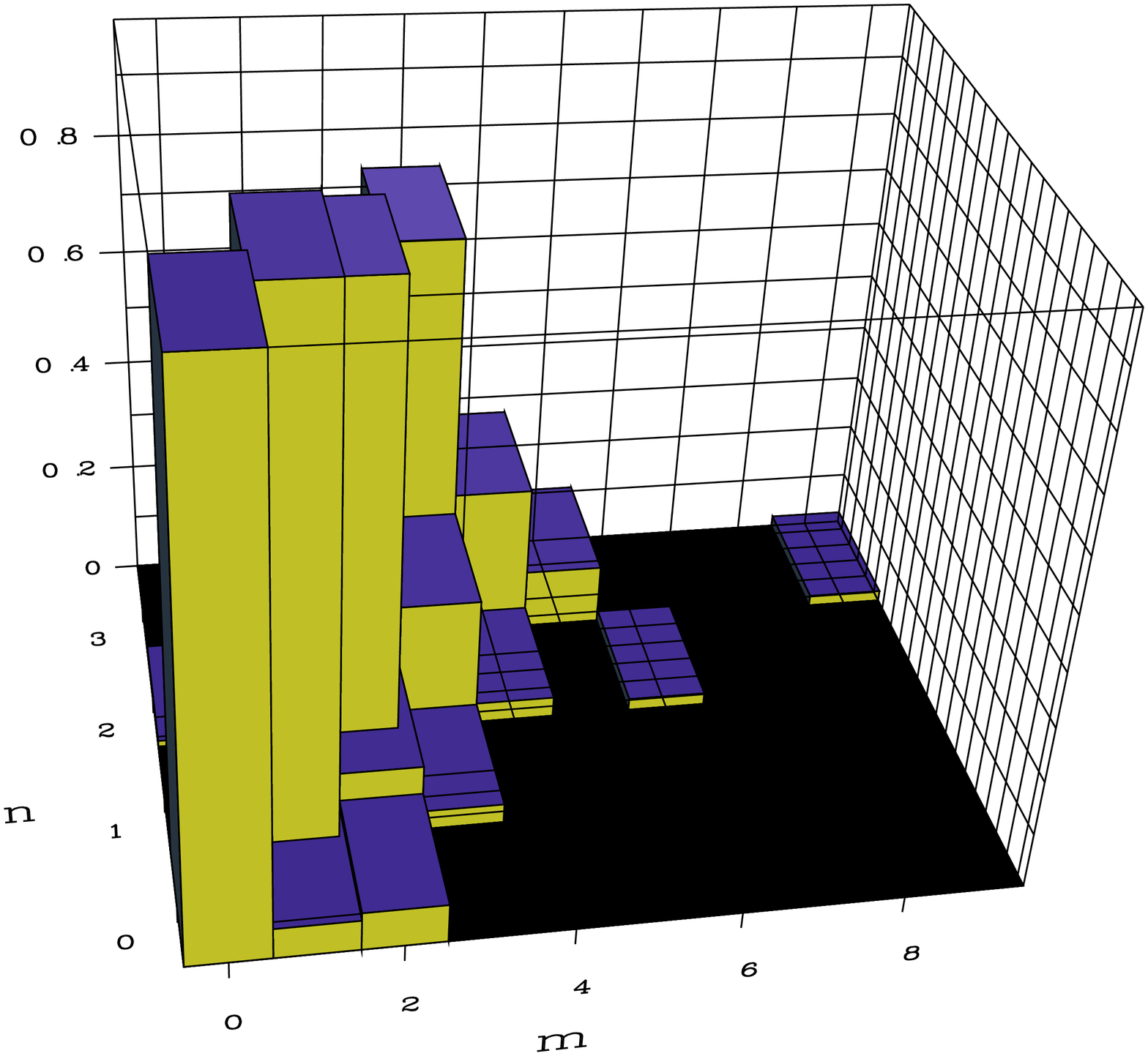}
\epsfxsize=.5
\hsize\leavevmode\epsffile{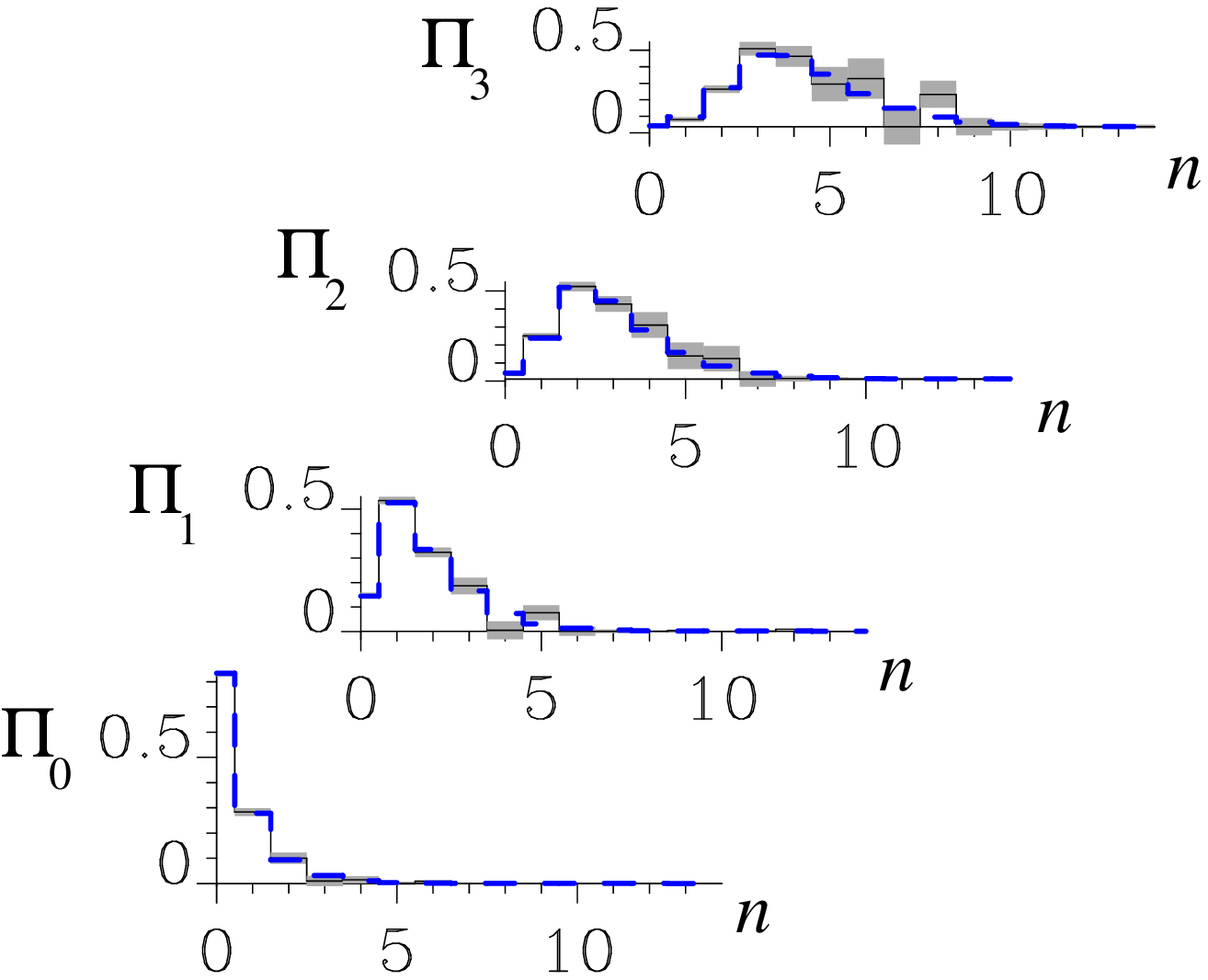}
\end{center}
\caption{Maximum likelihood reconstruction of the same POVM of
Fig.~\ref{f:tomog} with the same parameters, but here only $5\times
  10^4$ simulated homodyne measurements are employed. The statistical
  error bars are obtained by bootstrapping, i.e. by calculating the
  variance using the data of $50$ numerical experiments.  Notice that
  the result is statistically less noisy than the results presented in
  Fig.~\ref{f:tomog} even if here less measurements are employed:
  maximum likelihood is usually a better estimator.}
\labell{f:maxlik}\end{figure}

\section{History of quantum tomography}\labell{s:history}
In this section a brief historical perspective (see
also{\cite{bilkenttom,reviews}}) on quantum tomography is presented.
Already in 1957 Fano\cite{fano} stated the problem of quantum state
measurement, followed by rather extensive theoretical work. It was
only with the proposal by Vogel and Risken\cite{vogel}, however, that
homodyne tomography was born. The first experiments
followed\cite{raymer} by showing reconstructions of coherent and squeezed
states. The main idea at the basis of these works, is that it is
possible to extend to the quantum domain the algorithms that are
conventionally used in medical tomographic imaging to recover
two-dimensional distributions (say of mass) from unidimensional
projections in different directions.  However, these first tomographic
methods are unreliable for the measurement of unknown quantum states,
since some arbitrary smoothing parameters have to be introduced.

A new approach to optical tomography was then proposed\cite{dmp,dlp}
which allows to recover the quantum state of the field $\varrho$ (and
also the mean values of system operators) directly from the data,
abolishing all the sources of systematic errors. Only statistical
errors (that can be reduced arbitrarily by collecting more
experimental data) are left. Quantum tomography has been then
generalized to the estimation of arbitrary observable of the
field\cite{tokio}, to any number of modes\cite{homtom}, and to
arbitrary quantum systems via group theory\cite{grouptomo}, with
further improvements such as noise deconvolution\cite{deconv},
adaptive tomographic methods\cite{adaptive}, and the use of
max-likelihood strategies\cite{maxlik}, which has made possible to
reduce dramatically the number of experimental data, with negligible
bias for most practical cases of interest.  The latest developments
are based on a general method\cite{orth}, where the tomographic
reconstruction is based on the existence of spanning sets of
operators, of which group tomography\cite{grouptomo} is just a special
case.

\section*{Acknowledgments}
\addcontentsline{toc}{section}{Acknowledgements}

We acknowledge financial support by INFM PRA-2002-CLON and MIUR for
Cofinanziamento 2003 and ATESIT project IST-2000-29681.

\end{document}